\documentstyle[prd,aps]{revtex}

\begin{document}
\newcommand{\be}{\begin{eqnarray}}
\newcommand{\ee}{\end{eqnarray}}
\newcommand{\bi}{\bibitem}
\newcommand{\lar}{\leftarrow}
\newcommand{\rar}{\rightarrow}
\newcommand{\lrar}{\leftrightarrow}
\newcommand{\nc}{\newcommand}

\newcommand{\om}{\omega}
\newcommand{\ds}{\partial \!  \! \! /}
\newcommand{\Zs}{Z \! \! \! \! /}
\newcommand{\Ws}{W \! \! \! \! \! /}

\draft

\twocolumn[\hsize\textwidth\columnwidth\hsize\csname
@twocolumnfalse\endcsname

\title{ 
One more mechanism of electric charge non-conservation
}

\author{
Alexander D. Dolgov$^{(a,b)}$, 
Hideki Maeda$^{(c,d)}$ 
and Takashi Torii$^{(c)}$}

\address{ $^{(a)}$
INFN, sezione di Ferrara, via Paradiso, 12, 44100 - Ferrara,
Italy
}

\address{ $^{(b)}$
ITEP, Bol. Cheremushkinskaya 25, Moscow 113259, Russia
}

\address{ $^{(c)}$
Advanced Research Institute for Science and Engineering,
Waseda University, Tokyo 169-8555, Japan
}

\address{$^{(d)}$
Department of Physics, 
Waseda University, Tokyo 169-8555, Japan
}

\date{\today}

\maketitle

  
\begin{abstract}
If photon mass is not exactly zero, though arbitrary small, electric
charge would not be conserved in interaction of charged particles with
black holes. Manifestations of such mechanism of electric charge
non-conservation are discussed.
\end{abstract}


\vskip2pc]



Conservation of electric charge is an intrinsic property of the
standard massless electrodynamics because the Maxwell equations
\be
\nabla_\mu F^{\mu\nu} = J^\nu,
\label{maxwell}
\ee
automatically impose the condition of current conservation,
$\nabla_\mu J^\mu = 0$. To avoid this restriction one has to
change the theory prescribing a non-vanishing mass to photon. So
instead of the Maxwell equations the Proca equations for a massive
vector field should be used:
\be
\nabla_\mu F^{\mu\nu} + m^2 A^\nu = J^\nu.
\label{proca}
\ee
Now non-conservation of electric current may be compatible with the
equations of motion.

However such a model of electric charge non-conservation encounters
serious difficulties~\cite{okun78-z,okun78-v} because of very tight 
experimental (observational) upper limits on the photon
mass, $m< 10^{-33}$ eV, for a review see Refs.~\cite{mgamma}. 
If the photon mass is introduced simply by addition of $m^2A^2/2$-term
into Lagrangian (hard mass introduction) the theory 
with nonconserved current would be
nonrenormalizable, and the amplitudes in higher orders of perturbation
theory strongly rise with energy. Perturbative amplitude of
emission of longitudinal photons contains extremely large factors
$\sim (E/m)^n$, where $E$ is the characteristic energy of the process
with charge nonconservation, e.g. for electron decay, $e \rar \nu
+\gamma$ it is $E\sim 0.5$ MeV. As was argued in Ref.~\cite{okun78-v}
(see also \cite{tsypin89}) summation of radiative corrections to the
$e$-$\nu$-$\gamma$-vertex (or to any other charge nonconserving one) leads
to ``self-healing'' of the charge non-conservation: the probability of
any process where charge is not conserved becomes exponentially
small. 

If electromagnetic $U(1)$-symmetry is broken 
spontaneously~\cite{okun78-z,okun78-v} by the
Higgs mechanism (soft mass introduction) the theory would be
renormalizable and neither rising with energy amplitudes nor
self-healing would be present. In such a model, however, a light charged
scalar particle should exist in contradiction with experiment. The
contradiction would disappear if the charge of new light boson would
be very small at the level of $10^{-3}\, e$, where $e$ is the charge of
electron~\cite{millicharge}. 

A new idea was put forward recently in Ref.~\cite{dubovsky00}
based on the assumption that our four-dimensional world is a brane
embedded in the higher dimensional space with infinite extra 
dimensions~\cite{highD}. The authors of the papers~\cite{dubovsky00}
suggested that the charged
particles can leak into extra dimensions leaving impression that
electric charge is nonconserved in our four-dimensional world even
with strictly massless photons. According to this work, observation of
electric charge nonconservation could be a proof of existence of large
extra dimensions.

In this paper we would like to discuss a different mechanism of
electric charge nonconservation which, though demands a non-zero value
of the photon mass, could be efficient with an arbitrary small value
of the latter without leakage into higher dimensions. For example
this mechanism would operate if the Compton wave length of the photon
is equal to cosmological horizon and $m\sim H$, where $H$ is the
Hubble parameter, or even for much smaller values of the mass. The
arguments that the photon mass may be nonzero due to quantum effects
in de Sitter space-time were recently presented in
Ref.~\cite{prokopec02}. The mechanism is based on the well known fact
that black holes do not have any hairs related to massive vector
fields. Thus a charged particle captured by black hole would disappear
without leaving any trace in the form of the Coulomb field. The Coulomb field
begins to die down gradually when the charged particle approaches
horizon and the effect is nonvanishing for any small value of the photon
mass with discontinuous limit when $m\rar 0$~\cite{vilenkin}.

Let us consider spherically symmetric uniformly charged shell with 
radius $R_c$ 
concentric with a spherically symmetric matter shell with radius
$R_m$ whose mass is $M$. 
Let us assume for simplicity that the mass of the charged shell
is negligible in comparison with the mass of the matter shell, so the
gravitational field is completely determined by the matter shell, We
will always assume that the charged shell is outside the matter shell,
$R_m<R_c$
and calculate the electric field created by this
configuration.

Let us first consider the case when there is no black hole or, in other
words, the radius of the matter shell is larger than its gravitational
radius, $R_g<R_m$.
To obtain simple jump 
conditions for the Proca field $A_{\mu}$ across the matter shell, we adopt 
the coordinate system $(u,r,\theta,\phi)$ in which the 
metric is continuous;
\begin{eqnarray}
ds^2 = K^2du^2+2dudr-R(r)^2d\Omega^2,
\label{metricinside}
\end{eqnarray}
for the inner Minkowski region $R_m(1-K)<r<R_m$ and
\begin{eqnarray}
ds^2 = \Bigl(1-\frac{R_g}{r}\Bigr)du^2 +2dudr-r^2d\Omega^2, 
\label{metricoutside}
\end{eqnarray}
for the outer Schwarzschild region $r>R_m$.
Here $K = (1-R_g/R_m)^{1/2}$ and $R(r) = [r-R_m(1-K)]/K$ is the
area radius. In 
this coordinate system, both components $A_u$ and $A_r$ of 
vector potential are 
non-vanishing even in the static case. For $R_m(1-K)\le r<R_m$, the
coordinates $u$ and $r$ are  related to
the standard coordinates ${\bar u}$ and
${\bar r}$ by the  transformation
${\bar u}=Ku$ and ${\bar r}=R(r)$,
i.e., $r=R_m(1-K)$ corresponds to the physical center. 

Inside the matter shell, 
the Proca field satisfies the following equations:
\begin{eqnarray}
{1\over R^2}\left( R^2 A'_u \right)' - {m^2  \over K^2}\, A_u
=0, \quad A_u=K^2A_r,
\end{eqnarray}
where prime denotes derivative with respect to radial coordinate $r$. 
The solution which is regular at the center is
\begin{eqnarray}
A_u=C_3 \frac{\sinh m{\bar r}}{{\bar r}},
\end{eqnarray}
where $C_3$ is a constant determined from 
the junction conditions at the matter shell. 

Outside the matter shell, the Proca
field in the Schwarzschild background satisfies the following equations:
\begin{eqnarray}
{1\over r^2}\left( r^2 A'_u \right)' - {m^2 r \over r - R_g}\, A_u
=0,  
\label{eqout}
\end{eqnarray}
\begin{eqnarray}
A_r=\Bigl(1-\frac{R_g}{r}\Bigr)^{-1}A_u,
\label{eqout2}
\end{eqnarray}
where prime denotes derivative with respect to radial coordinate $r$. 
It is convenient to introduce the new
function $q = r\,A_u $ and the new variable $y = 2m(r-R_g)$. In terms of
these quantities the Proca equation (\ref{eqout}) takes the form:
\begin{eqnarray}
\frac{d^2q}{dy^2} - \left( {1\over 4} + {\mu \over 2y}\right)\, q =0,
\label{q''2}
\end{eqnarray}
where $\mu := m R_g$. This equation has the form of the Whittaker 
differential equation and the solution is expressed
in terms of confluent
hypergeometric functions~\cite{gr}:
\begin{eqnarray}
q(y) &=& C_i\,y\, e^{-y/2}\,\Phi (1+\mu/2,2,y)
\nonumber \\
&&
+B_i\,y\, e^{-y/2}\,\Psi (1+\mu/2,2,y) ,
\label{y-sol2}
\end{eqnarray}
where the coefficients $C_i$ and $B_i$ are constants to be determined
from the boundary conditions and the junction conditions. They have different
values in different space regions and this is indicated by the index $i$. $i=1$
implies the  outside region of the charged shell, while $i=2$ does inside. When
$y\to \infty$, the function $\Phi$ rises as
$y^{-1+\mu/2}\exp (y)$, while $\Psi$ decreases as $\Psi \sim
y^{-(1+\mu/2)}$. So for $r>R_c$ the solution is given by the expression
(\ref{y-sol2}) with $C_1 =0$, while for $r<R_c$ both coefficients $B_2$
and $C_2$ may be non-zero.

Since there are two shells, we should consider the junction conditions of
the Proca field there. 
There is no charge distribution
on the matter shell $r=R_m$, hence the Proca field and its
first  derivative are continuous there, i.e., 
$[A_u]^+_-=0$ and $[A'_u]^+_-=0$, where $[f(r)]^+_-=f(r+0)-f(r-0)$. On the
other hand,  the junction condition on the charged shell is the usual one
that the discontinuity of the derivative of the electric field on the shell is
equal to the charge density: 
\be
[A_u]^+_-=0, \;\;\;\; [A'_u]^+_- = 4\pi \sigma.
\label{deltaE}
\ee 
%
Now all the coefficients $B_i$ and $C_i$ can be determined.
We obtain the coefficient of the outer region 1
from the junction conditions as
\begin{eqnarray}
B_1&=&\frac{Qe^{y_c/2}(F\Phi_c-G\Psi_c)}
{2my_cR_cF(\Phi_c\Psi'_c-\Phi'_c\Psi_c)},
\label{shellb1}
\end{eqnarray}
where $Q = 4\pi \sigma R_c^2$ is the total charge of the shell, prime denotes derivative with respect to $y$ and
\begin{eqnarray}
F&=&2mR_m y_m\Psi'_m-P\Psi_m,
\\
G&=&2mR_m y_m\Phi'_m-P\Phi_m,
\\
P&=&
\frac{y_m}{K}\left(\frac{mR_m}{\tanh(mR_m)}-1\right)
-2mR_g+mR_m y_m.
\nonumber
\\
\end{eqnarray}
The sub-index $m$ and $c$ imply the values evaluated at 
$y=y_m =2m(R_m-R_g)$ (i.e., $r=R_m$) and 
$y=y_c =2m(R_c-R_g)$ (i.e., $r=R_c$), respectively.

{}From now on we will consider some limiting cases.
The confluent hypergeometric functions behaves around  $y=0$ as
\begin{eqnarray}
\Phi(1+\mu/2,2,y)&\sim&1+\left(\frac12+\frac{\mu}{4}\right)y,\\
\Psi(1+\mu/2,2,y)&\sim&\Gamma(1+\mu/2)^{-1}\frac{1}{y}+\Gamma(\mu/2)^{-1}\ln y \nonumber \\
 &&+ O(1),
\label{Psi0}
\end{eqnarray}
where $\Gamma$ is gamma function. It should be noted that the $y\to 0$ limit 
is realized by different two cases, one of which is $r\to R_g$ and
the other is $mr\to 0$. In the latter case, $O(1)$ term in Eq.~(\ref{Psi0})
vanished since $\Psi(1,2,y)=1/y$ becomes exact. This fact will be used to 
calculate to next leading order below.

First we examine the limit of small photon mass, i.e.
$mR_{c,m,g}\to 0$ and $mr\to 0$.
Taking this limit, we find
\begin{eqnarray}
B_1&=&-Q(1-mR_g-\mu\gamma/2),
\end{eqnarray}
where $\gamma$ is Euler's constant and this implies
\begin{eqnarray}
A_u \sim -\frac{Q}{r}(1-mr).
\label{sol-1}
\end{eqnarray}
This is the standard expression for electrostatic Coulomb potential in massless
electrodynamics. In
this case the limit to zero mass of the photon is smooth.

Now let us consider a different limiting situation when $R_m \to R_g$,
which corresponds to the space-time with a black hole. We find that
\begin{eqnarray}
B_1 &\sim& 
\frac{Q\exp(y_c/2)\Phi_c}{2my_cR_c(\Phi_c\Psi'_c-\Phi'_c\Psi_c)}
\nonumber \\
&&\times
\Biggl[1+\frac{2mR_g^{1/2}\Gamma(1+\mu/2)\Psi_c}{\Phi_c}
\left(\frac{mR_g}{\tanh(mR_g)}-1\right)^{-1}
\nonumber \\
&& 
\;\;\;\;\; \;\;\times (R_m-R_g)^{1/2}\Biggr].
\end{eqnarray}
We will obtain essentially the same solution if we assume that the
black hole is already formed, i.e., we take $R_m < R_g$ from the very
beginning. In this case the solution for the vector potential inside
the charged shell should not contain the function $\Psi$ because the 
solution would be singular at the horizon, $\Psi \sim y^{-1}$,
and the electric field $E =-\nabla A_t$ would be infinite for 
$y\rar 0$. The conditions of continuity of the potential at $r=R_c$ 
and the discontinuity of its derivative (\ref{deltaE}) permit to
determine both coefficients $B_1$ and $C_2$.

Next we make the charged shell close to the gravitational radius,
$R_c \to R_g$, keeping the mass $m$ non-zero.
The result is
\begin{eqnarray}
B_1 \sim -Q\frac{R_c-R_g}{R_g}\Gamma(1+\mu/2).
\end{eqnarray}
Hence the electric field vanishes as the charged shell approaches
the gravitational radius and  disappears completely when
the charged shell is swallowed by the black hole. This indicates the
non-conservation of the charge.

We can consider another limiting case: $R_m \to R_g$ and after that
$m\to 0$. In this case the electrostatic potential
takes the form
\begin{eqnarray}
A_t \sim -{Q\over r}\,{R_c-R_g \over R_c}[1-m(r-R_g)],
\label{sol-2}
\end{eqnarray}
for relatively small distances, $mr\ll1$.
There is a finite change of the potential discontinuous at the point
$m=0$~\cite{vilenkin}.


To see how this discontinuity arises is convenient to present the
solution in the (realistic) case of small photon mass such that the 
product $m R_{c,m,g} \ll 1$ and $mr \ll 1$ but keeping the finite
difference $(R_m-R_g)$. The effective charge $q$ in this case is
\begin{eqnarray}
q= Q\,\left[1-{R_m \over R_c}\,{ (mR_m)^2 \sqrt{{R_m / (R_m-R_g)}}
\over 3 + (mR_m)^2 \sqrt{{R_m /( R_m-R_g)}}} \right]
\label{qeff}
\end{eqnarray}
One can easily see from this expression how the two limiting cases
presented above are realized. The boundary between them is determined 
by the value of the product:
\begin{eqnarray}
\Delta = (mR_m)^2 \sqrt{ R_m /(R_m-R_g)}.
\label{delta}
\end{eqnarray}
If $\Delta \ll 1$ the usual massless electrodynamics with conserved 
eclectic charge would be valid. On the other hand, if 
$R_m-R_g \ll R_m(R_m m)^4$, charge non-conservation would be effective.

The solutions that we have found above are true in asymptotically flat
space-time. So there may be essential corrections to them at the
distances comparable to cosmological horizon. Still the main result
that electric charge must be non-conserved in the process of black
hole capture of charged particles, if photon has non-zero but even 
vanishingly small mass, most probably remains true. 
After the charged particle capture the Coulomb field that existed at
large distances must also disappear. One can obtain a feeling how it
disappears in adiabatic approximation assuming that the charged sphere
slowly approaches the gravitational radius, $R_c = R_c (t)\rar R_g$. In
this process the Coulomb field should be radiated away in
the form of longitudinal photons. To calculate the law of their
emission one has to solve time-dependent problem when both $A_t$ and
$A_r$ are non-vanishing. Technically it is much more difficult than
considered here adiabatic approach and we will consider it in future
work. 

According to these considerations charged
particle capture by  a black hole could lead to a cosmological electric
charge asymmetry and electric forces at the scales below 
Compton wave length of photon, 
$r<1/m$ might be cosmologically interesting. They even could lead to
an accelerated expansion. Electric charge asymmetry might be also
generated by black hole evaporation in way analogous to generation of
baryon asymmetry by the similar process of black hole 
evaporation~\cite{bh-bs}. Dynamics and efficiency of generation 
of electric charge asymmetry, however, would be different from 
generation of baryonic one 
because of presence of long range forces in the former case.
 
There could be also some interesting processes in particle physics due
to interactions with virtual black holes, for example the decay
$e \rar \nu +$ photons, inequality of proton and electron charges and
some other effects which have already been discussed in the literature 
in connection with the hypothesis of electric charge nonconservation
(see the papers~\cite{okun78-z,okun78-v,millicharge}). 
It is difficult, however, to estimate the probability of
such processes because calculations of the effects associated with
virtual black hole interactions demand still missing knowledge of
strong quantum gravity. Electric charge nonconservation may be
possibly observed at high energy collidors if the fundamental scale of 
gravity is much lower than the usual Planck scale~\cite{arkani98}. 
One may expect in this case that small size black holes would be
abundantly produced in next generation accelerators or in high energy 
cosmic rays~\cite{bh-prod} (see however criticism of
ref.~\cite{voloshin01}) and the probability of accompanying processes
with nonconserved electric charge would be quite high.

\section*{Acknowledgments}
A.D. is grateful to Yukawa Institute for Theoretical Physics for
hospitality during period when this work was done, and to I.B. Khriplovich
for discussions.


\end{document}